\documentstyle[11pt,mrs2001]{article}
\pdfoutput=1
\begin{document}

\title{Reflections on the direct detection of particle dark matter}

\author{R.H. Sanders} 

\affil{Kapteyn Astronomical Institute, University of Groningen,
9700 AV Groningen, The Netherlands}

\begin{abstract}
The LUX experimental group has just announced the most stringent upper limits
so far obtained on the cross section of 
WIMP-nucleon elastic scattering \cite{lux1}.  
This result is a factor
of two to five below the previous best upper limit \cite{xen1} and effectively
rules out earlier suggestions of low mass WIMP detection signals.
The experimental expertise exhibited by this group is
extremely impressive, but the fact of continued negative results
raises the more basic question of
whether or not this is the right approach to solving the
dark matter problem.  
Here I comment upon this question, using as a basis 
the final chapter
of my book on dark matter \cite{sanbook},
somewhat revised and extended.
I muse on dark matter and its alternative, modified Newtonian
dynamics, or MOND.

\end{abstract}

In the context of the present cosmological paradigm,
less than 5\% of the world is potentially visible,
and even 70\% of this normal baryonic component is not detected.  
The remainder of the mass-energy content of the Universe
is thought to consist partly of dark matter that is unidentified, and
primarily of dark energy \index{dark energy} of even more uncertain nature.
The dark matter fills the Universe, promotes structure formation\index{structure formation, gravitational collapse}
and accounts for the discrepancy between the visible and
dynamical mass of bound astronomical systems such as galaxies
and clusters; it is the major constituent of such systems.
In order to cluster on the scale of galaxies at sufficiently
early epochs, the dark matter must be essentially
pressureless, i.e., non-relativistic at the time it
decoupled from photons\index{photons} and other particles.  The dark energy\index{dark energy},
which may be identified with the zero-point energy of
the vacuum, causes the present accelerated expansion of
the Universe and provides the 70\% contribution to the
density budget for the flat Universe required by
observations of CMB \index{CMB, Cosmic Microwave Background} anisotropies.

So the two primary constituents of the world are
detected only by their dynamical effects:  dark energy \index{dark energy}
which affects in an observable way the expansion history
of the Universe, and dark matter that clumps and dominates
the mass budget of bound gravitating systems.
Some members of our scientific community are uncomfortable 
with the concept of dark energy\index{dark energy}.  There is a theoretical
problem with its small magnitude (on the scale of
particle physics).  Moreover, in so far as this peculiar 
fluid can be represented by a cosmological constant\index{cosmological constant}, it
does not dilute with the expansion of the Universe as does
the density of ordinary matter.
Why, then, are the
densities of these two substances so nearly comparable 
at the present epoch?  Why are we now witness to this
remarkable coincidence?  If the dark energy\index{dark energy} is dynamical and 
identified with
some new field, should we not see other manifestations of
that field, such as fifth force effects, evident as
a violation of the universality of free fall?  

Dark matter is more comprehensible and appealing to two scientific communities:
physicists and astronomers.
Theoretical physicists like it because the best bet extension
of their Standard Model\index{Standard Model of particle physics}, Supersymmetry\index{Supersymmetry}, provides candidate
particles with, possibly, the right properties.   
Many very competent experimentalists are willing
to spend a significant fraction of their lives looking for it
in the laboratory, even though its nature, and therefore
the detection strategy, remains uncertain.  The search for dark matter
has become a large industry with all the vested interests
of a large industry; this is because detection would be one
of the major scientific discoveries of all time.  
The odds are long but the stakes are high. 
However, it should be realized that non-detection is not falsification\index{falsification}, even in the context
of a particular class of dark matter candidates, superpartners
for example.  And given that the range of possible candidates is limited
only by the human imagination, then unsuccessful dark matter
searches can always be accommodated in the context of the paradigm.

Astronomers like dark matter.  If you cannot see it, you
can use it to produce rotation curves of any sort, stabilize disks, make warps,
promote mergers, explain anomalous lensing, make large scale
structure... It is fun to
simulate because all you need is Newtonian gravity\index{Newtonian gravity} which
is easy to compute.  Again, the one thing you cannot do
is falsify it; almost any astronomical observation can be accommodated
by dark matter particularly if one ignores the systematics of galaxy rotation curves and near
perfect global scaling relations. These result
from poorly understood ``gastrophysics" -- gas dynamical process, star
formation, feedback.  This of course is accompanied by a large leap
of faith that someday these processes will be understood and
all will be explained.

While these two communities, physicists and astronomers, have
found a common interest in dark matter, we should remember
that they are, in fact, rather distinct communities with
different methodologies and different criteria for interpretation
of results.  Physicists are more prepared to go beyond
known physics (Supersymmetry\index{Supersymmetry}, after all, is an extension of
known physics) while astronomers are more conservative in
this respect (and well they should be; interpretation of
astronomical results can otherwise become quite bizarre).
Physicists know that galaxy rotation curves are flat and
that this constitutes a primary evidence for dark matter
that should manifest itself locally.  They do not know about, and are
not much interested in,
the regularities of rotation curves or global scaling
relations; these are details for astronomers.  Astronomers know,
because physicists tell them, that particle dark matter is 
well-motivated and that it is proper to invoke dark matter in
understanding astronomical observations.  Of course, this is 
simplistic; there are individuals who move easily between both communities,
have a broader overview, and strongly support the concept
of dark matter.  The point is that both physicists and astronomers
find it useful to invoke dark matter from their own different
vantage points, but I argue that from both sides 
the dark matter concept is fundamentally not
falsifiable.

It was Karl Popper \cite{popp} who first emphasized the importance of
falsification\index{falsification} in eliminating scientific theories and
progressing to new ideas.  This surely must be true,
given the inherent asymmetry between falsification and
verification.  Of course,
to be falsified a theory must be in practice falsifiable; this would
seem to be a hallmark of good theory (for Popper\index{Popper, K.}, a theory
that is not falsifiable is not scientific).  Dark matter as a theory
misses this attribute (which is not to say that it is wrong).
In my opinion the most serious challenge for the dark matter 
hypothesis is the
existence of an algorithm -- MOND\index{MOND, MOdified Newtonian Dynamics} -- that can predict the form
of rotation curves from the observed distribution of detectable
matter.  This is something that dark matter does not naturally permit
because it is a different sort of fluid and not subject to
all of the physical effects that influence baryonic matter and
its distribution.  Moreover, MOND\index{MOND, MOdified Newtonian Dynamics}, as a theory, is inherently 
falsifiable.  If
particles with the right properties to constitute the cold dark
matter are found tomorrow, then MOND\index{MOND, MOdified Newtonian Dynamics} is out of the window.
In that sense it is a better theory (which is not to say that 
it is right). 

With respect to progress through falsification\index{falsification}, 
the reality is never so simple and certainly is not in this
case.  In the issue of dark
matter vs. MOND\index{MOND, MOdified Newtonian Dynamics} we are not dealing just with two theories
but with two competing paradigms in the sense meant by Thomas Kuhn
\cite{kuhn}.
Thirty-five years ago it was becoming generally recognized that something
was missing in large astronomical systems like galaxies
and clusters.  This recognition was not an instantaneous
process.  When an observation
runs counter to our expectation, we do not always recognize 
the anomaly; recall the early attempts to fit observed galaxy rotation curves
exhibiting no evidence for a decreasing rotation velocity
by models having a built-in Keplerian decline.
But by 1980, it was no longer in doubt that {\it something} 
had been discovered;  but {\it what} that something is, in fact, has
never been so certain.  Dark matter was the initial, and natural,
first attempt at a solution to this astronomical anomaly.
With respect to galaxies, the concept of dark halos\index{dark halo}
was already in place as a means of taming the instability
of rotationally supported systems.  At the same time,
it became appreciated that the difficulty of forming the 
observed structure in an expanding Universe with a finite 
lifetime could be overcome by adding a universal 
non-baryonic matter component.  And at the same time,
particle physics seemed to be providing a host of particle
candidates.  The astronomical anomaly, the cosmological
necessity, and the particle physics possibility combined to give
the dark matter hypothesis the status of a paradigm:  a framework,
a set of assumptions that are not questioned, a list of
problems that are to be addressed as well as problems
that are not to be addressed.

Not long after the discovery of the anomaly (1983), the hypothesis
of modified Newtonian dynamics (MOND\index{MOND, MOdified Newtonian Dynamics}) emerged, and this
proposal can be clearly associated with a single individual -- 
Moti Milgrom \cite{m83a} (there were other such ideas in circulation, but none
of them successfully addressed so many aspects of the phenomena).
At the time MOND\index{MOND, MOdified Newtonian Dynamics} was what Kuhn\index{Kuhn, T.S.} would call an ``anticipation"
and not a response to a crisis with dark matter -- there was
no such crisis.  MOND\index{MOND, MOdified Newtonian Dynamics} was truly an alternative to the dark 
matter hypothesis -- in fact, the only viable alternative explanation for
the observed anomaly.  But while the dark matter hypothesis
attracted a large following early on -- thanks primarily to its range
of application, from galaxies to cosmology -- MOND\index{MOND, MOdified Newtonian Dynamics} languished
for some years with only a handful of advocates (myself included).
But now due to its proven predictive power, at least on the scale of galaxies,
the development of a reasonable relativistic extension (thanks
primarily to Jacob Bekenstein \cite{bek04}), and simple frustration with the
absence of dark matter particle detection, MOND\index{MOND, MOdified Newtonian Dynamics}
has also achieved the status of a competing paradigm,
although one still supported by a small minority of the relevant
communities.
 
Supporters of different paradigms give different weight
to different experimental or observational facts, and this
makes the issue of falsification\index{falsification} rather murky.  For example,
supporters of the dark matter paradigm tend to emphasize 
cosmological aspects.  They would argue that on a cosmological 
scale General Relativity\index{General Relativity} with dark matter
(and dark energy\index{dark energy}) presents a coherent picture.  The observed
phenomenology of the CMB\index{CMB, Cosmic Microwave Background} 
fluctuations\index{fluctuations} and the formation
and distribution of galaxies
on large scale is explained in the context of the Concordance Cosmology.  
They tend to dismiss galaxy scale phenomenology and its systematics
as being essentially due to messy baryonic physics which will
someday be understood in the context of the larger picture.
MOND\index{MOND, MOdified Newtonian Dynamics} supporters, on the other hand, emphasize the 
regularities in galaxy phenomena: the predicted appearance
of a discrepancy in low surface brightness\index{surface brightness: low, high} systems, the
near perfect Tully-Fisher law (the baryonic mass-rotation velocity
relation), the ability of the algorithm
to predict the amplitude and form of rotation curves (including
details).  They are rather 
dismissive of the cosmological evidence, at least until
the recent development of the relativistic extension.

The point is (and this is essentially Kuhn\index{Kuhn, T.S.}'s point) arguments
between supporters of different paradigms are somewhat akin
to arguments about religion.  The assumptions and the
criteria for truth are different.  Most scientists do not
feel the need to adopt a new paradigm unless the old one
is in crisis, so we may ask:  Is dark matter in crisis?
Are there fundamentally un-resolvable anomalies within the
context of dark matter?  Again, most supporters would 
answer in the negative (but they would then, wouldn't they?).
I personally think that there is a crisis -- not only the
observational difficulties of dark matter on galaxy scales 
(see discussion below) but also a creeping
crisis provoked by the non-detection of dark matter particles.
I have argued that this is not properly a falsification\index{falsification}, but
it surely must be a worry.  At what point will experimentalists
stop searching for these elusive particles and shift to activities
more likely to produce positive results?  At what point will theorists
tire of more and more speculative conjectures on the nature
of hypothetical undetectable matter?  And what if apparent
deviations from Newtonian gravity\index{Newtonian gravity} or dynamics are seen
in the Solar System\index{Solar System}? 
 
It is certainly true that, for scientists taken
as a social group, most effort goes into
attempting to prove or strengthen the existing paradigm rather than
to challenge it.  Consistencies are
valued over anomalies, and uncomfortable facts are overlooked or
pushed into the category of complicated problems for the future.
This is probably necessary because normal science is a social
process and takes place in the context of a paradigm. 
The social phenomenon is reinforced by external considerations:
by competition for academic positions, by the necessity of
obtaining research grants.  I expect that it has
always been this way, but in a general sense (and
this is why Kuhn\index{Kuhn, T.S.} emphasizes the significance of ``scientific revolutions")
progress is a dialectic process and due to the
conflict of ideas rather than ``concordance".  By ``progress"
here I mean moving in a direction of increased understanding
of the world around us. Kuhn\index{Kuhn, T.S.} would certainly not agree
with this definition nor even with the
concept of progress as movement toward a goal, but I believe 
that it is meaningful.

With respect to MOND\index{MOND, MOdified Newtonian Dynamics}, I, and others, have been impressed because 
it explains and unifies aspects of galaxy phenomenology
which would appear to be disconnected in the context of
dark matter. The primary reason for MOND's staying power 
is its remarkable predictive
power on the scale of galaxies -- a predictive power that is not,
or cannot be, matched by dark matter as it is presently perceived.
Herein lies 
a minimalist definition of MOND -- a definition which is as free
as possible from emotive charge of concepts such as 
modified inertia or modified gravity:

\noindent {\it MOND is an algorithm that permits one to calculate the
distribution of force in an object from the observed distribution
of baryonic matter with only one additional universal constant having
units of acceleration.}

And it works!  The algorithm works very well 
on the scale of galaxies as evidenced
by the MOND determinations of the rotation curves of
spiral galaxies from the observed distributions of stellar
and gaseous mass.  This is quite problematic for Cold Dark Matter (CDM), 
because -- and I repeat -- it is not something that dark matter, as it is perceived to
be, can naturally do.  Moreover, MOND
explains or subsumes systematic aspects of galaxy photometry
and kinematics as a consequence of fundamental dynamical principles.
CDM can only address tight observational correlations, such as
the Tully-Fisher relation, 
as emerging from the process of
galaxy formation -- a process that by its nature is quite
haphazard, each
galaxy having its own unique history of formation, interaction,
and evolution.  It is difficult to imagine that the
ratio of baryonic to dark mass would be a constant in galaxies,
or even vary systematically with galaxy mass.  And yet this is
required, in a very precise way, to explain the baryonic Tully-Fisher
relation \cite{mcg} -- an exact correlation between the baryonic mass and the
asymptotic rotation velocity which supposedly is a property
of the dark matter halo.
Any initial intrinsic
velocity-mass relation of proto-galaxies would surely be erased in
the very stochastic processes of galaxy formation.  To believe that
poorly understood mechanisms such as ``feedback'' or ``self-regulation'' can
restore even tighter correlations demonstrates a remarkably naive acceptance
of the paradigm.
 
The phenomenological successes of MOND constitute a severe
challenge for CDM or any non-interacting 
dissipationless dark matter that 
clusters on the scale of galaxies -- in fact, some might even
say that it is a falsification of dark matter as it is 
perceived to be.  At the very least one can say that the
astronomical evidence for particle dark matter in galaxies,
such as the Milky Way, is far less firm than 
particle physicists have been
led to believe.  And where does this leave direct detection
experiments.  I certainly would not encourage anyone involved to
give up on such experiments; indeed, it is fortunate that
a number of experiments are underway -- experiments by
independent groups using different techniques.  This serves
as a useful cross-check on marginal or controversial 
claims, and the spin-off technology may well find other
applications.  But I doubt that this is the primary motive of
the experimenters in spite of the positive spin placed
on each new, more stringent upper limit.  Patience is 
not unlimited, nor is funding, and I would advise any young
physicist confronted by career choices to keep this
in mind.  Perhaps it is true that the emperor has no
clothes.

\vfill

\end{document}